\documentclass[pra,floatfix,twocolumn,superscriptaddress,showpacs]{revtex4}
\usepackage{amsmath}
\usepackage{graphicx}
\usepackage{amssymb}
\usepackage{epsfig}
\usepackage{amsfonts}

\def\extra#1{{}}
\def\>{\rangle}

\input{tcilatex}
\begin{document}

\title{Spin squeezing and concurrence}
\author{Xiaolei Yin}
\affiliation{Zhejiang Institute of Modern Physics, Department of Physics, Zhejiang
University, Hangzhou 310027, China}
\author{Xiaoqian Wang}
\affiliation{Zhejiang Institute of Modern Physics, Department of Physics, Zhejiang
University, Hangzhou 310027, China}
\affiliation{Department of Physics, Changchun University of Science and Technology,
Changchun 130022, China }
\author{Jian Ma}
\affiliation{Zhejiang Institute of Modern Physics, Department of Physics, Zhejiang
University, Hangzhou 310027, China}
\author{Xiaoguang Wang}
\email{xgwang@zimp.zju.edu.cn}
\affiliation{Zhejiang Institute of Modern Physics, Department of Physics, Zhejiang
University, Hangzhou 310027, China}
\keywords{spin squeezing,concurrence,entanglement}
\pacs{03.65.Ud,03.67.2a}

\begin{abstract}
We study the relations between spin squeezing and concurrence, and find that
they are qualitatively equivalent for an ensemble of spin-1/2 particles with
exchange symmetry and parity, if we adopt the spin squeezing criterion given
by the recent work (G. T\'{o}th \textit{et al}. Phys. Rev. Lett. \textbf{99}%
, 250405 (2007)). This suggests that the spin squeezing has more
intimate relations with pairwise entanglement other than
multipartite entanglement. We exemplify the result by considering a
superposition of two Dicke states.
\end{abstract}

\maketitle

\section{Introduction}

As an important resource of quantum information and computation,
entanglement~\cite{Einstein,Schrodinger} has attracted much attention in
recent years~\cite%
{Bennett1993,Bennett1992,Ekert1991,Wang2001,WangSolomon,Pan2001,Vidal2003,VidalPalacios,Leibfried,Andre}%
. How to measure and detect entanglement is crucial for both theoretical
investigations and potential practical applications~\cite%
{Bennett1996,Vedral1997}.\ The entanglement of a two-qubit system can be
well quantified by the concurrence~\cite{Wootters1997,Wootters1998}.
However, quantification of many-body entanglement is still an open question
in quantum information.

It is well known that there are close relations between entanglement and
spin squeezing~\cite%
{SorensenMolmer,Sorrensen2001,Kitagawa1993,Wineland1994,Kitagawa2001,WangSangders2003,Jin2007,Jafarpour}%
. There are several definitions of spin squeezing parameters~\cite%
{Sorrensen2001,Kitagawa1993,Wineland1994}, which are studied in different
papers. The squeezing parameter $\xi _{R}^{2}$ defined by Wineland \textit{%
et al. }is closely related to multipartite entanglement. It has been proven
that~\cite{Sorrensen2001}, for an ensemble of spin-1/2 particles, if this
squeezing parameter is less than one, the state is entangled. The advantages
of spin squeezing parameters in detecting entanglement have been shown in
both theoretical and experimental aspects.

The squeezing parameter $\xi _{S}^{2}$ defined by Kitagawa and Ueda is
relevant to pairwise entanglement~\cite{Kitagawa2001}. And for states with
exchange symmetry and parity, a simple quantitative relation between $\xi
_{S}^{2}$\ and concurrence was given~\cite{WangSangders2003}. Furthermore,
it has been shown that the spin squeezing and pairwise entanglement are
equivalent for states generated by the one-axis twisting Hamiltonian~\cite%
{WangSangders2003}. However, even for states with a fixed parity, such as
the states generated by one-axis twisting Hamiltonian with a transverse
field, $\xi _{S}^{2}$ is not always equivalent to concurrence~\cite{Wang2004}%
. Inspired by recent works~\cite{Toth2007,Toth2009}, where a set of
generalized spin squeezing inequalities are developed, one can define
another spin squeezing parameter $\xi _{T}^{2}$ from one of the inequalities~%
\cite{Wang2009}. Similar to parameter $\xi _{R}^{2}$, one advantage of the
parameter $\xi _{T}^{2}$ is that we can firmly say that the state is
entangled if $\xi _{T}^{2}<1$. However, if parameter $\xi _{S}^{2}<1 $, we
cannot say the state is entangled, although this parameter is closely
related to entanglement.

Reference~\cite{Kitagawa2001} found that spin squeezing according to
parameter $\xi _{S}^{2}$ is equivalent to the minimal pairwise correlation $%
\mathcal{C}_{\vec{n}_{\perp },\vec{n}_{\perp }}$ along the direction $\vec{n}%
_{\perp }$ (which is perpendicular to the mean spin direction) for symmetric
states. It was further found~\cite{Xiaoqian10} that for the symmetric
states, the spin squeezing defined via $\xi _{T}^{2}$ is equivalent to
minimal pairwise correlation $\mathcal{C}_{\vec{n},\vec{n}}$ along an
arbitrary direction $\vec{n}$. For states with a fixed parity, the relations
between the two parameters $\xi _{S}^{2}$ and $\xi _{T}^{2}$ are more
evident. It will be seen from Sec.~3 that $\xi _{T}^{2}$ contains the term $%
\xi _{S}^{2}$, and the spin squeezing results from just the competition
between pairwise correlation along the direction $\vec{n}_{\perp }$ and that
along the mean spin direction. So, in this sense, the parameter $\xi
_{T}^{2} $ is a natural generalization of $\xi _{S}^{2}$.

We find that for states with exchange symmetry and parity, the spin
squeezing parameter $\xi _{T}^{2}$ is qualitatively equivalent to the
concurrence in characterizing pairwise entanglement. In other words, the
spin squeezing parameter and concurrence emerge and vanish at the same time.
This finding is of significance to entanglement detection in experiments. As
we all know, entanglement detectors such as entropy and concurrence are
generally not easy to measure, especially for physical systems like BEC, for
which we cannot address individual atoms. However, spin squeezing parameters
are relatively easy to measure in experiments, since they only consist of
expectations and variances of collective angular momentum operators.
Nevertheless, the traditional spin squeezing parameter $\xi _{S}^{2}$ is not
always equivalent to concurrence even for states with exchange symmetry and
parity. As $\xi _{T}^{2}$ is qualitatively equivalent to concurrence for an
ensemble of spin-1/2 particles with exchange symmetry and parity, it is
better than $\xi _{S}^{2}$\ in detecting pairwise entanglement.

The paper is organized as follows: In Sec.~II, we give the definitions of
the two spin squeezing parameters and concurrence. In Sec.~III, we give the
concrete forms of the spin squeezing parameters and the concurrence for
states with exchange symmetry and parity. The relations between these two
parameters and concurrence were given in Sec.~IV. We exemplify the result by
considering superpositions of Dicke states in Sec.~V. Finally, Sec.~VI is
devoted to conclusion.

\section{Spin squeezing parameters and concurrence}

To study spin squeezing, we consider an ensemble of $N$ spin-1/2 particles.
For the sake of describing many-particle systems, we use the total angular
momentum operators%
\begin{equation}
J_{\alpha }=\frac{1}{2}\sum_{k=1}^{N}\sigma _{k\alpha },~~~\left( \alpha
=x,y,z\right) ,  \label{collective}
\end{equation}%
where $\sigma _{k\alpha }$ are the Pauli matrices for the $k$-th spin. Now,
we give the definitions of the two spin squeezing parameters. The first one
is defined as~\cite{Kitagawa1993},
\begin{equation}
\xi _{S}^{2}=\frac{4\min (\Delta J_{\vec{n}_{\perp }})^{2}}{N},
\label{kitagawa}
\end{equation}%
where $\vec{n}_{\perp }$ is perpendicular to the mean spin direction $\vec{n}%
=\frac{\langle \vec{J}\rangle }{|\langle \vec{J}\rangle |}$. Since the
system has the exchange symmetry, the total angular momentum is $j=\frac{N}{2%
}$. For spin coherent states~\cite{Kitagawa1993}, $\Delta J_{\vec{n}_{\perp
}}=\frac{j}{2}$, and $\xi _{S}^{2}=1$. In the following, we consider states
with exchange symmetry.

The next spin squeezing parameter is based on the generalized spin squeezing
inequalities given by T\'{o}th \textit{et al}.~\cite{Toth2009}, and is
defined as~\cite{Wang2009}%
\begin{equation}
\xi _{T}^{2}=\frac{\lambda _{\min }}{\langle \vec{J}^{2}\rangle -\frac{N}{2}}%
,  \label{Toth}
\end{equation}%
where $\lambda _{\min }$ is the minimum eigenvalue of
\begin{equation}
\Gamma =(N-1)\gamma +G  \label{gamma}
\end{equation}%
with $G_{kl}=\frac{1}{2}\left\langle J_{k}J_{l}+J_{l}J_{k}\right\rangle $, $%
(k,l=x,y,z)$ the correlation matrix, and $\gamma _{kl}=G_{kl}-\left\langle
J_{k}\right\rangle \left\langle J_{l}\right\rangle $ the covariance matrix.
For our states, $\langle \vec{J}^{2}\rangle -\frac{N}{2}=j\left( j+1\right)
-j=j^{2}$.

The two-qubit entanglement is quantified by the concurrence, whose
definition is given by~\cite{Wootters1998}
\begin{equation}
C=\max \{\lambda _{1}-\lambda _{2}-\lambda _{3}-\lambda _{4},0\},
\label{concurrence_one}
\end{equation}%
where $\lambda _{1}\geq \lambda _{2}\geq \lambda _{3}\geq \lambda _{4}$ are
the square roots of eigenvalues of $\tilde{\rho}\rho $. Here $\rho $ is the
reduced density matrix of the system, and%
\begin{equation}
\tilde{\rho}=(\sigma _{y}\otimes \sigma _{y})\rho ^{\ast }(\sigma
_{y}\otimes \sigma _{y}),  \label{roub}
\end{equation}%
where $\rho ^{\ast }$ is the conjugate of $\rho $. If $C>0$, the system
displays pairwise entanglement.

\section{States with parity}

To study the relations between spin squeezing parameters and concurrence, we
consider a class of states with even (odd) parity, which means a state in
the $(2j+1)$-dimensional space with only even (odd) excitations of spins.
These kinds of states are widely studied, e.g., the states generated by the
one-axis twisting model~\cite{Kitagawa1993}. The states with even parity
possess important properties, $\left\langle J_{\alpha }\right\rangle =0$, $%
\left\langle J_{\alpha }J_{z}\right\rangle =\left\langle J_{z}J_{\alpha
}\right\rangle =0$, $\alpha =x,y$, which means the mean spin direction is
along the $z$-axis, and the covariances between $J_{z}$ and $J_{\alpha }$
are zero. Thus, equation~(\ref{gamma}) becomes
\begin{equation}
\Gamma =\left(
\begin{array}{ccc}
N\left\langle J_{x}^{2}\right\rangle & \frac{N\left\langle \left[ J_{x},J_{y}%
\right] _{+}\right\rangle }{2} & 0 \\
\frac{N\left\langle \left[ J_{x},J_{y}\right] _{+}\right\rangle }{2} &
N\left\langle J_{y}^{2}\right\rangle & 0 \\
0 & 0 & N(\Delta J_{z})^{2}+\langle J_{z}\rangle ^{2}%
\end{array}%
\right) ,  \label{correlation matrix}
\end{equation}%
where $\left[ A,B\right] _{+}=AB+BA$, and equation~(\ref{Toth}) reduces to~%
\cite{Wang2009}
\begin{equation}
\xi _{T}^{2}=\min \left\{ \xi _{S}^{2},\varsigma ^{2}\right\} ,
\label{Toth_}
\end{equation}%
where
\begin{eqnarray}
\varsigma ^{2} &=&\frac{4}{N^{2}}\left[ N(\Delta J_{z})^{2}+\langle
J_{z}\rangle ^{2}\right]  \notag \\
&=&1+\left( N-1\right) \left( \left\langle \sigma _{1z}\sigma
_{2z}\right\rangle -\left\langle \sigma _{1z}\right\rangle ^{2}\right)
\notag \\
&=&1+(N-1)C_{zz},  \label{z_derection}
\end{eqnarray}%
with $C_{zz}$ the two-spin correlation function along $z$ direction. The
explicit form of $\xi _{S}^{2}$ could be obtained as~\cite{WangSangders2003}%
\begin{eqnarray}
\xi _{S}^{2} &=&\frac{2}{N}\left( \langle J_{x}^{2}+J_{y}^{2}\rangle
-|\langle J_{-}^{2}\rangle |\right)  \notag \\
&=&1-2\left( N-1\right)  \notag \\
&&\times \left[ \left\vert \langle \sigma _{1-}\sigma _{2-}\rangle
\right\vert -\frac{1}{4}\left( 1-\left\langle \sigma _{1z}\sigma
_{2z}\right\rangle \right) \right] ,  \label{ketagawa-general}
\end{eqnarray}%
where we have used the following relations%
%=======================================================================
\begin{eqnarray}
\left\langle J_{\alpha }\right\rangle &=&\frac{N}{2}\left\langle \sigma
_{1\alpha }\right\rangle ,  \notag \\
\langle J_{\alpha }^{2}\rangle &=&\frac{N}{4}+\frac{N(N-1)}{4}\langle \sigma
_{1\alpha }\sigma _{2\alpha }\rangle ,  \notag \\
\langle J_{-}^{2}\rangle &=&N(N-1)\langle \sigma _{1-}\sigma _{2-}\rangle ,
\label{collective-local}
\end{eqnarray}%
which connect the local expectations with collective ones.

For such states, the significance of $\xi _{S}^{2}$ and $\xi _{T}^{2}$ and
the relations between them are clear. According to the parameter $\xi
_{S}^{2}$, a state is squeezed when the minimum variance of angular momentum
in the $\vec{n}_{\perp }$-plane is smaller than $\frac{j}{2}$, while
according to $\xi _{T}^{2}$, the variance in the mean spin direction $\vec{n}
$ is also considered. {Equation~(\ref{z_derection}) represents the pairwise
correlation along the mean spin direction, and this can be viewed as a
complement of }$\xi _{S}^{2}${, which only considers squeezing in the }$\vec{%
n}_{\perp }$-plane\textit{.} {Thus, }$\xi _{T}^{2}$ {can be regarded as a
generalization of }$\xi _{S}^{2}${, and }when $\xi _{S}^{2}<\varsigma ^{2}$,
the parameter $\xi _{T}^{2}$ reduces to $\xi _{S}^{2}$.

To calculate concurrence, we first need to calculate the two-body reduced
density matrix, which can be written as~\cite{WangSangders2003}%
\begin{equation}
\rho =\left(
\begin{array}{cccc}
v_{+} & 0 & 0 & u^{\ast } \\
0 & y & y & 0 \\
0 & y & y & 0 \\
u & 0 & 0 & v_{-}%
\end{array}%
\right) ,  \label{rho}
\end{equation}%
where%
%=======================================================================
\begin{eqnarray}
v_{\pm } &=&\frac{1}{4}\left( 1\pm 2\langle \sigma _{1z}\rangle +\langle
\sigma _{1z}\sigma _{2z}\rangle \right) ,  \notag \\
u &=&\langle \sigma _{1-}\sigma _{2-}\rangle ,~~~y=\frac{1}{4}\left(
1-\left\langle \sigma _{1z}\sigma _{2z}\right\rangle \right) ,
\label{rho-relations}
\end{eqnarray}%
in the basis $\left\{ \left\vert 00\right\rangle ,\left\vert 01\right\rangle
,\left\vert 10\right\rangle ,\left\vert 11\right\rangle \right\} $. Then the
concurrence is given by
\begin{equation}
C=2\max \left\{ 0,~|u|-y,~y-\sqrt{v_{+}v_{-}}\right\} .  \label{conc}
\end{equation}

One key observation is that
\begin{equation}
y^{2}-v_{+}v_{-}=-\frac{1}{4}C_{zz}.
\end{equation}%
Thus,
\begin{equation}
\varsigma ^{2}=1-4(N-1)(y+\sqrt{v_{+}v_{-}})(y-\sqrt{v_{+}v_{-}}).
\end{equation}%
From equations~(\ref{z_derection}),~(\ref{ketagawa-general}), and~(\ref%
{rho-relations}), we obtain%
\begin{eqnarray}
\xi _{S}^{2} &=&1-2(N-1)\left( \left\vert u\right\vert -y\right) ,  \notag \\
\xi _{T}^{2} &=&\min \{1-2(N-1)\left( \left\vert u\right\vert -y\right) ,
\notag \\
&&1-4(N-1)(y+\sqrt{v_{+}v_{-}})(y-\sqrt{v_{+}v_{-}})\}.  \label{xi_conc}
\end{eqnarray}%
Now, one can see that the squeezing parameters are related to the
concurrence shown in equation~(\ref{conc}). The relations between $\xi
_{S}^{2}$ and $C$ have been studied~\cite{WangSangders2003}. In the
following, we consider the squeezing parameter $\xi _{T}^{2}$, and prove
that it is qualitatively equivalent to the concurrence in detecting pairwise
entanglement.

\section{Relations between spin squeezing parameters and concurrence}

Firstly, we prove that for a state with exchange symmetry and parity, if
concurrence $C>0$, it must be spin squeezed according to the criterion $\xi
_{T}^{2}<1$. From equation~(\ref{conc}) we note that when $C>0$, there are
two cases, $C=\left\vert u\right\vert -y>0$ or $C=y-\sqrt{v_{+}v_{-}}>0$.
However, since the density matrix $\rho $ is positive, we find $\sqrt{%
v_{+}v_{-}}\geq \left\vert u\right\vert $, then immediately
\begin{equation}
\left( |u|-y\right) \left( y-\sqrt{v_{+}v_{-}}\right) \leq 0,
\label{positive}
\end{equation}%
which means $\left\vert u\right\vert -y$ and $y-\sqrt{v_{+}v_{-}}$ cannot be
positive\ simultaneously. Therefore, if $C>0$, we have~\cite{Vidal2006}%
\begin{equation}
C=\left\{
\begin{array}{ll}
2\left( \left\vert u\right\vert -y\right) ,~~~ & \left\vert u\right\vert >y,
\\
2\left( y-\sqrt{v_{+}v_{-}}\right) ,~~~ & y>\sqrt{v_{+}v_{-}}.%
\end{array}%
\right.
\end{equation}%
According to equations~(\ref{Toth_}) and~(\ref{xi_conc}), we get the
following relations%
\begin{equation}
\xi _{T}^{2}=\left\{
\begin{array}{ll}
1-\left( N-1\right) C,~~~ & \left\vert u\right\vert >y, \\
1-2\left( N-1\right) \left( y+\sqrt{v_{+}v_{-}}\right) C,~~~ & y>\sqrt{%
v_{+}v_{-}},%
\end{array}%
\right.
\end{equation}%
since $C>0$, there always be $\xi _{T}^{2}<1$.

Now, we prove that if the state is spin squeezed $\left( \xi
_{T}^{2}<1\right) $, concurrence $C>0$. If $\xi _{T}^{2}<1$, there are two
cases, $\xi _{T}^{2}=\xi _{S}^{2}<1$ or $\xi _{T}^{2}=\varsigma ^{2}<1$. As
discussed above, according to equations~(\ref{xi_conc}) and~(\ref{positive}%
), $\xi _{S}^{2}<1$ and $\varsigma ^{2}<1$ could not occur simultaneously.
Therefore, if $\xi _{T}^{2}=\xi _{S}^{2}<1$, we have~\cite{Vidal2004}%
\begin{equation}
C=\frac{1-\xi _{T}^{2}}{N-1},
\end{equation}%
while if $\xi _{T}^{2}=\varsigma ^{2}<1$, we have%
\begin{equation}
C=\frac{1-\xi _{T}^{2}}{2\left( N-1\right) \left( y+\sqrt{v_{+}v_{-}}\right)
}.
\end{equation}%
Therefore, if the state is squeezed, concurrence $C>0$.

\begin{table*}[tbp]
\caption{Spin squeezing parameters and concurrence for states with exchange
symmetry and parity.}
\label{tab_squeezing_conc}
\begin{center}
\begin{tabular}{c||l|l|c}
\hline
\parbox{2 cm} {\vspace{2mm}~\vspace{2mm}} & \multicolumn{2}{c|}{Pairwise
entangled ($C>0$)} & \parbox{2 cm}{\vspace{2mm} Unentangled\vspace{2mm}} \\
\hline
\parbox{2 cm} {Concurrence} &
\parbox{4 cm}{\vspace{2mm}$C=2(\left \vert
u\right \vert -y)>0$\vspace{2mm}} & $C=2(y-\sqrt{v_{+}v_{-}})>0$ & $C=0$ \\
\hline
$\xi _{S}^{2}$ &
\parbox{4
cm}{\vspace{2.2mm}$\xi_{S}^{2}=1-(N-1)C<1$\vspace{2mm}} & $\xi _{S}^{2}>1$ &
$\xi _{S}^{2}\geq 1$ \\ \hline
$\xi _{T}^{2}$ &
\parbox{5
cm}{\vspace{2.2mm}$\xi_{T}^{2}=1-\left( N-1\right) C<1$\vspace{2mm}} & %
\parbox{6 cm}{$\xi_{T}^{2}=1-2(N-1)(y+\sqrt{v_{+}v_{-}}) \times {C<1}$} & $%
\xi _{T}^{2}\geq 1$ \\ \hline
\end{tabular}%
\end{center}
\end{table*}

The relations between spin squeezing and concurrence is displayed in Table %
\ref{tab_squeezing_conc}, and we can see that, for a symmetric state, $%
\xi_{T}^{2}<1$ is qualitatively equivalent to $C>0$, that means spin
squeezing according to $\xi_{T}^{2}$ is equivalent to pairwise entanglement.
Although $\xi_{S}^{2}<1$ indicates $C>0$, when $C=2(y-\sqrt {v_{+}v_{-}})>0$%
, we find $\xi_{S}^{2}>1$. Therefore, a spin-squeezed state ($\xi_{S}^{2}<1$%
) is pairwise entangled, while a pairwise entangled state may not be
spin-squeezed according to the squeezing parameter $\xi_{S}^{2}$. Then, we
come to the conclusion that for states with exchange symmetry and parity,
the spin squeezing parameter $\xi _{T}^{2}$ is qualitatively equivalent to
the concurrence in characterizing pairwise entanglement. In the following,
we will give some examples and applications of our result.

\section{Examples and Applications}

We first consider a superposition of Dicke states with parity, and then
consider states without a fixed parity. The states under consideration are
all based on Dicke states~\cite{Dicke1954}, and are defined as%
\begin{equation}
|n\rangle _{N}\equiv |\frac{N}{2},-\frac{N}{2}+n\rangle ,~~~n=0,\ldots ,N,
\label{Dickestate}
\end{equation}%
where $|0\rangle _{N}\equiv |\frac{N}{2},-\frac{N}{2}\rangle $ denotes a
state for which all spins are in the ground states, and $n$ is the
excitation number of spins. Such states are elementary in atomic physics,
and may be conditionally prepared in experiments with quantum non-demolition
technique~\cite{Molmer1998,Mandel,Lemer2009}.

As we consider the state with even parity, we choose a simple superposition
of Dicke states as%
\begin{equation}
|\psi _{D}\rangle =\cos \theta |n\rangle _{N}+e^{i\varphi }\sin \theta
|n+2\rangle _{N},~~~n=0,\ldots ,N-2  \label{Dicke State_e}
\end{equation}%
with the angle $\theta \in \lbrack 0,\pi )$ and the relative phase $\varphi
\in \lbrack 0,2\pi )$. We can easily check that, for the superposition state
in equation~(\ref{Dicke State_e}), the mean spin direction is along the $z$%
-axis. The expressions for the relevant spin expectation values can be
obtained as%
\begin{eqnarray}
\left\langle J_{z}\right\rangle &=&m+2\sin ^{2}\theta ,  \notag \\
\langle J_{z}^{2}\rangle &=&m^{2}+\left( 4m+1\right) \sin ^{2}\theta ,
\notag \\
\langle J_{+}^{2}\rangle &=&\langle J_{-}^{2}\rangle =\frac{1}{2}e^{i\varphi
}\sin 2\theta \sqrt{\mu _{n}},  \label{mean_of_J}
\end{eqnarray}%
where $m=n-\frac{N}{2},$ and $\mu _{n}=\left( n+1\right) \left( n+2\right)
\left( N-n\right) \left( N-n-1\right) $.

By substituting equations~(\ref{mean_of_J}) to equation~(\ref{z_derection})
and~(\ref{ketagawa-general}), it is easy to get
\begin{eqnarray}
\xi _{S}^{2} &=&1-\frac{2}{N}\{\left\vert \sin \theta \cos \theta
\right\vert \sqrt{\mu _{n}}  \notag \\
&&-4[m^{2}+4(m+1)\sin ^{2}\theta ]-N^{2}\}  \label{xi-theta}
\end{eqnarray}%
and
\begin{eqnarray}
\varsigma ^{2} &=&\frac{4}{N}\left[ m^{2}+4\left( m+1\right) \sin ^{2}\theta %
\right]   \notag \\
&&-\frac{4(N-1)}{N^{2}}\left[ m+2\sin ^{2}\theta \right] ^{2}.
\end{eqnarray}%
From the results in~\cite{Wang2002} we can easily get~\cite{Vidal2006}%
\begin{eqnarray}
u &=&\frac{e^{i\varphi }\sin 2\theta }{2N(N-1)}\sqrt{\mu _{n}},  \notag \\
y &=&\frac{N}{4(N-1)}-\frac{[m^{2}+4(m+1)\sin ^{2}\theta ]}{N(N-1)},  \notag
\\
\sqrt{v_{+}v_{-}} &=&\frac{\sqrt{(N^{2}-2N+4\left\langle
J_{z}^{2}\right\rangle )^{2}-16(N-1)^{2}\left\langle J_{z}\right\rangle ^{2}}%
}{4N(N-1)}.  \notag \\
&&  \label{rho_results}
\end{eqnarray}%
Insert equation~(\ref{rho_results}) to equation~(\ref{conc}), one can get
the expression of concurrence.

\begin{figure}[ptb]
\begin{center}
\includegraphics[
height=2.2217in, width=2.9456in ]{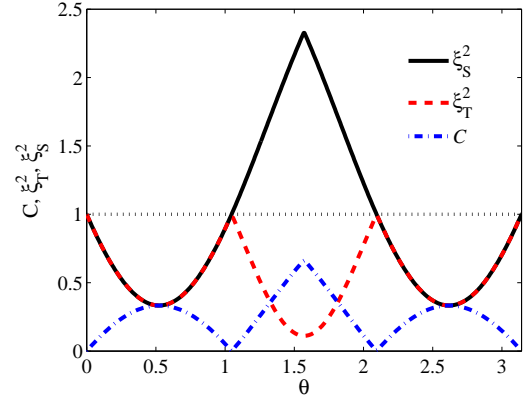}
\end{center}
\caption{Spin squeezing parameters $\protect\xi_{S}^{2}$ and $\protect\xi%
_{T}^{2}$, and concurrence as functions of $\protect\theta$ for $N=3$ and $%
n=0.$}
\label{fig1}
\end{figure}
%End Expansion

\begin{figure}[tbp]
\begin{center}
\includegraphics[
height=2.2217in, width=2.9456in ]{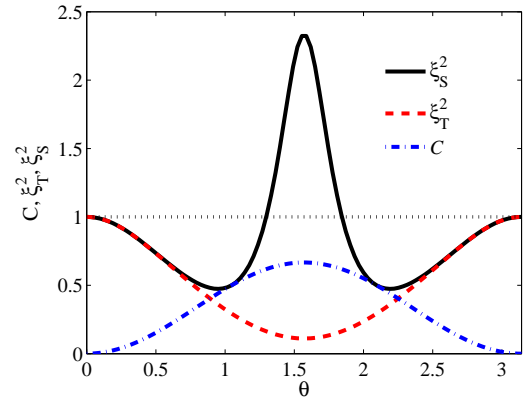}
\end{center}
\caption{Spin squeezing parameters $\protect\xi _{S}^{2}$ and $\protect\xi %
_{T}^{2}$, and concurrence as functions of $\protect\theta $ for $N=3$ and $%
n=0.$ The numerical result give $\protect\xi _{S}^{2}=7/3$, $\protect\xi %
_{T}^{2}=1/9$, and $C=2/3$, when $\protect\theta =\protect\pi /2.$ }
\label{fig2}
\end{figure}

In figure~\ref{fig1}, we plot these two spin squeezing parameters and
concurrence versus $\theta $ in one period. We observe that for $\theta \in
(0,\pi /3)\cup (2\pi /3,\pi )$, $\xi _{T}^{2}=\xi _{S}^{2}<1$, therefore the
state is spin squeezed in the $x$-$y$ plane, moreover, as $C>0$, the state
is pairwise entangled. For $\theta \in (\pi /3,2\pi /3)$, it is obviously
that the state is also pairwise entangled, since $C>0,$ while spin squeezing
occurs in the $z$-axis since $\xi _{T}^{2}<1$ while $\xi _{S}^{2}>1$. The
results show clearly that $\xi _{T}^{2}<1$ is equivalent to $C>0$. But if we
adopt $\xi _{S}^{2}<1$ as squeezing parameter, the spin squeezing is not
qualitatively equivalent to concurrence.

The equivalence of $\xi _{T}^{2}<1$ and $C>0$ for states with parity has
been demonstrated above. Here, we discuss states without parity to see the
relations between spin squeezing and entanglement. For simplicity, we choose%
\begin{equation}
|\psi _{D}\rangle =\cos \theta |n\rangle _{N}+e^{i\varphi }\sin \theta
|n+1\rangle _{N},~~~n=0,\ldots ,N-1.  \label{Dick-state-n+1}
\end{equation}%
Specifically, if $\theta =\frac{\pi }{2}$, $n=0$ or $n=N-2$, the above state
degenerates to the W state. Moreover, when $N=3$, equation~(\ref%
{Dick-state-n+1}) reduces to
\begin{equation}
|\psi _{D}\rangle =\frac{1}{\sqrt{3}}(|110\rangle +|101\rangle +|011\rangle
).
\end{equation}%
The two-qubit reduced density matrix becomes%
\begin{equation}
\rho =\frac{1}{3}\left(
\begin{array}{cccc}
0 & 0 & 0 & 0 \\
0 & 1 & 1 & 0 \\
0 & 1 & 1 & 0 \\
0 & 0 & 0 & 1%
\end{array}%
\right) ,
\end{equation}%
and using equation~(\ref{conc}) we find $C=\frac{2}{3}$. We can also get the
expectations of spin components, $\left\langle J_{z}\right\rangle =-\frac{1}{%
2}$, $\left\langle J_{x}^{2}\right\rangle =\left\langle
J_{y}^{2}\right\rangle =\frac{7}{4}$, $\left\langle J_{y}^{2}\right\rangle =%
\frac{1}{4}$, and then we can easily get the spin squeezing parameters, $\xi
_{S}^{2}=\frac{7}{3}$ and $\xi _{T}^{2}=\frac{1}{9}$. The numerical results
for $\xi _{T}^{2}$ is displayed in figure~\ref{fig2}, which coincide with
the special result. It is interesting to see that, although $|\psi
_{D}\rangle $ has no parity, the state is entangled $\left( C>0\right) $ and
is spin squeezed according to $\xi _{T}^{2}$ in the entire interval.
However, according to parameter $\xi _{S}^{2}$ the state is not squeezed in
the middle region. Therefore, we find that $\xi _{T}^{2}$ is more effective
than $\xi _{S}^{2}$ in detecting pairwise entanglement.

\section{\textbf{Conclusion}}

{In conclusion, we have studied the relations between spin squeezing and
pairwise entanglement. We have considered two types of spin squeezing
parameters }$\xi _{S}^{2}$\textit{\ and }$\xi _{T}^{2}$ {, and the pairwise
entanglement is characterized by concurrence }$C${. We find that, for states
with exchange symmetry and parity, spin squeezing according to }$\xi
_{T}^{2} ${\ is qualitatively equivalent to pairwise entanglement. In
detecting pairwise entanglement, parameter $\xi _{T}^{2}$ is more effective
than parameter $\xi _{S}^{2}$. }

It is important to emphasize that, the above conclusion can be extended to
the states without (even or odd) parity. For states with properties $%
\left\langle J_{\alpha }\right\rangle =0$, $\left\langle J_{\alpha
}J_{z}\right\rangle =\left\langle J_{z}J_{\alpha }\right\rangle =0$, $\alpha
=x,y$, we can have the same conclusion that spin squeezing and pairwise
entanglement are qualitatively equivalent. The following superposition of
Dicke states are examples: $\left\vert \psi _{D^{\prime }}\right\rangle
=\cos \theta |n\rangle _{N}+e^{i\varphi }\sin \theta |n+n^{\prime }\rangle
_{N},$ $n=0,\ldots ,N-n^{\prime }$, for all $n^{\prime }\geq 3$. As we have
seen, parameter $\xi _{S}^{2}$ is a key factor in $\xi _{T}^{2}$ for our
states. The present results imply that the spin squeezing has more intimate
relations with pairwise entanglement.

\section*{Acknowledgements}

This work is supported by NSFC with grant No. 10874151 and 10935010; and the
Fundamental Research Funds for the Central Universities.

\end{document}